\documentclass[prd,preprint,superscriptaddress,preprintnumbers,byrevtex,showpacs]{revtex4}

\usepackage{epsfig}
\usepackage{isolatin1}

\begin{document} 

\title{Testing Color Evaporation in Photon-Photon Production of $J/\psi$  at
  CERN LEP II}

\author{O.\ J.\ P.\ \'Eboli}
\email{eboli@fma.if.usp.br}
\affiliation{Instituto de F\'{\i}sica, 
Universidade de S\~ao Paulo, S\~ao Paulo -- SP, Brazil.}

\author{E.\ M.\ Gregores}
\email{gregores@ift.unesp.br}
\affiliation{Instituto de F\'{\i}sica Te\'orica, 
Universidade Estadual Paulista, S\~ao Paulo -- SP, Brazil.}

\author{J.\ K.\ Mizukoshi}
\email{mizuka@fma.if.usp.br}
\affiliation{Instituto de F\'{\i}sica, 
Universidade de S\~ao Paulo, S\~ao Paulo -- SP, Brazil.}

%%%%%%%%%%%%%%%%%%%%%%%%%%%%%%%%%%%%%%%%%%%%%%%%%%%%%%%
\begin{abstract} 
  
The DELPHI Collaboration has recently reported the measurement of
$J/\psi$ production in photon-photon collisions at LEP II. These newly
available data provide an additional proof of the importance of colored
$c \bar{c}$ pairs for the production of charmonium because these data
can only be explained by considering resolved photon processes. We
show here that the inclusion of color octet contributions to the
$J/\psi$ production in the framework of the color evaporation model is
able to reproduce this data. In particular, the transverse-momentum
distribution of the $J/\psi$ mesons is well described by this model.

\end{abstract}

\preprint{
%%%\textbf{DRAFT 04}}
\preprint{IFT-P.034-03}
%\preprint{MADPH-02-xxx}
\preprint{IFUSP-1577/2003}}


\maketitle

%%%%%%%%%%%%%%%%%%%%%%%%%%%%%%%%%%%%%%%%%%%%%%%%%%%%%%%%%%%%%%%%%%%%%%
\section{Introduction}

DELPHI Collaboration recently released preliminary measurements of the
transverse momentum spectrum of $J/\psi$ mesons produced in $\gamma \gamma$
collisions at LEP \cite{Todorova-Nova:2001pt,Abdallah:2003du}. This new data allow
further tests of models for charmonium production. We show here that the Color
Evaporation Model (CEM) reproduces these new results using the same single
non-perturbative parameter that has been obtained from previous analysis of
charmonium photo-- and hadro--production.  These newly available data provide
an additional proof of the importance of colored $c \bar{c}$ pairs for the
production of charmonium, as the data on this region can only be explained by
considering resolved photon processes, which forms colored $c \bar{c}$ pairs
in the leading order. The CEM for charmonium production incorporates these
colored pairs into the total yield of charmonium in a very simple and
economical way.

The Tevatron data \cite{cdf,d0} on charmonium production at high $p_T$ changed
the way we understand charmonium production. The presently successful models
are based in two key considerations: i) onium production is a two--step
process where a heavy quark pair is produced first, followed by the
non--perturbative formation of the asymptotic states, and ii) color octet as
well as singlet $c\bar{c}$ states contribute to the production of charmonia.
These features are incorporated in the Non-Relativistic QCD (NRQCD)
factorization approach \cite{Bodwin:1994jh, Braaten:1996pv}, in the Color
Evaporation Model \cite{Amundson:1995em, Amundson:1996qr,Eboli:1996gd}, and in
the Soft Color Interaction Model \cite{Edin:1997zb}.

The Color Evaporation Model simply states that charmonium production is
described by the same dynamics as $D \overline{D}$ production, {\em i.e.}, by
the formation of a $c\bar{c}$ pair in any color configuration. Rather than
imposing that the $c\bar{c}$ pair is in a color singlet state in the short
distance perturbative processes, it is argued that the appearance of color
singlet asymptotic states solely depends on the outcome of non-perturbative
large distance fluctuations of quarks and gluons.  These large distance
fluctuations are considered to be complex enough for the occupation of
different color states to approximately respect statistical counting.  In
fact, it is indeed hard to imagine that a color singlet state formed at a
range $m_{\psi}^{-1}$ would survive to form a $\psi$ at a range
$\Lambda_{QCD}^{-1}$. Although far more restrictive than other proposals, CEM
successfully accommodates all features of charmonium production
\cite{Mariotto:2001sv, Kramer:2001hh, Schuler:1996ku}.

The CEM predicts that the sum of the production cross sections of all onium
and open charm states is described by
\begin{equation} 
\sigma_{\rm onium} = \frac{1}{9}\int_{2 m_c}^{2 m_D} 
dM_{c \bar{c}}~ \frac{d \sigma_{c \bar{c}}}{dM_{c\bar{c}}} \; , 
\label{sig:on} 
\end{equation} 
and 
\begin{equation} 
\sigma_{\rm open} = \frac{8}{9}  \int_{2 m_c}^{2 m_D} 
dM_{c\bar{c}}~\frac{d \sigma_{c\bar{c}}}{d M_{c \bar{c}}} 
+ \int_{2 m_D} dM_{c\bar{c}}~\frac{d\sigma_{c\bar{c}}}{dM_{c\bar{c}}} \; , 
\label{sig:op} 
\end{equation} 
where $M_{c\bar{c}}$ is the invariant mass of the $c\bar c$ pair. The factor
$1/9$ stands for the probability that a pair of charm quarks formed at a
typical time scale $1/M_\psi$ ends up as a color singlet state after
exchanging an uncountable number of soft gluons with the reaction remnants;
for further details see \cite{Amundson:1995em}.  One attractive feature of
this model is the relation between the production of charmonium and open
charm, which allows us to use the open charm data to normalize the
perturbative QCD calculation, and consequently to constrain the CEM
predictions.

Up to this point, the model has no free parameter in addition to the usual QCD
ones. In order to predict the production rate of a particular charmonium
state, let us say a $J/\psi$ meson, we must also know the fraction $\rho_\psi$
of produced onium states that materialize as this state ($J/\psi$),
\begin{equation}
    \sigma_\psi = \rho_\psi~\sigma_{\rm onium} \; .
\label{sig:psi}
\end{equation}
In its simplest version, the CEM assumes that $\rho_\psi$ is energy and
process independent, which is in agreement with the low energy measurements
\cite{Gavai:1994in,Schuler:1994hy}. Notice that $\rho_\psi$ is the solely free
parameter of the CEM, making this a very restrictive framework.  From the
charmonium photo--production, we determined that $\rho_\psi=0.43$--0.5
\cite{Amundson:1996qr}, a value that can be accounted for by statistical
counting of final states \cite{Edin:1997zb}.  The fact that all $\psi$
production data are described in terms of this single parameter, fixed by
$J/\psi$ photo--production, leads to parameter free predictions for $Z$-boson
decay rate into $\psi$ \cite{Gregores:1996ek}, and to charmonium production
cross section at Tevatron \cite{Eboli:1999hh} and HERA
\cite{Eboli:1998xx,Eboli:2003fr}, as well as in neutrino initiated reactions
\cite{Eboli:2001hc}.

%%%%%%%%%%%%%%%%%%%%%%%%%%%%%%%%%%%%%%%%%%%%%%%%%%%%%%%%%%%%%%%%%%%%%%
\section{Results}

The differential cross section for the inclusive process $e^+ e^- \to e^+ e^-
\gamma \gamma \to J/\psi X$ is
\begin{equation} 
\frac{d^2\sigma}{dp_T^2} = \sum_{A,B}
\int\!\!\!\int\!\!\!\int\!\!\!\int dy^+ dy^- dx_A dx_B
f_{\gamma/e^+}(y^+) f_{\gamma/e^-}(y^-) 
F_{A/\gamma}(x_A) F_{B/\gamma}(x_B)
\frac{d^2\hat\sigma(A B \to \psi Y)}{dp_T^2}  \; , 
\end{equation} 
where $f_{\gamma/e^\pm}$ is the bremsstrahlung photon distribution from an
electron/positron. We denoted the parton distribution function of the
photon by $F_{A[B]/\gamma}(x_{A[B]})$, where $x_{A[B]}$ is the fraction of the
photon momentum carried by the parton $A[B]$. For direct photon interactions
($A[B]\equiv\gamma$), we have $F_{A[B] / \gamma} (x_{A[B]}) =
\delta(x_{A[B]}-1)$.  We considered an average electron--positron
center--of--mass energy $2E_e=197$ GeV.  We also applied the experimental
$J/\psi$ rapidity cut $-2 < \eta_\psi < 2$, and imposed that the $\gamma
\gamma$ center--of--mass energy satisfies $W_{\gamma\gamma}<35$ GeV, where
$W_{\gamma\gamma}=2E_e\sqrt{y^+y^-}$.

In our calculation, we employed the Weiz\"acker-Williams approximation for the
photon distribution
\begin{equation}
f_{\gamma/e^\pm}(y) =\frac{\alpha_{em}}{2\pi} 
\left[ \frac{1+(1-y)^2}{y} \log\left(\frac{Q^2_{max}}{Q^2_{min}}\right) 
+ 2m_e^2 y \left( \frac{1}{Q_{max}^2} - \frac{1}{Q_{min}^2}\right) \right] \; ,
\end{equation}
with $Q^2_{min} = m_e^2 y^2/(1-y)$, and $Q^2_{max} = (E_e\theta)^2
(1-y) + Q^2_{min}$. Here, the fraction of the parent $e^\pm$ energy
($E_e$) carried by the photons is $y \; (= E_\gamma/E_e)$, and $\theta$
is the angular cut that guarantees that the photons are real. We used
$\theta = 0.032$ radians, as determined by the experiment.

The inclusive subprocess cross section $\hat\sigma(A B \to \psi Y)$ was
calculated using the CEM; see Eqs.\ (\ref{sig:on}) and
(\ref{sig:psi}).  The partonic subprocesses contributing to $J/\psi$
production are depicted in the Table \ref{process}. Notice that both
direct and resolved photons contribute to charmonium production in the
CEM.  We evaluated numerically the tree level helicity amplitudes of
the subprocesses displayed in Table \ref{process} using MADGRAPH
\cite{Stelzer:1994ta} and HELAS \cite{Murayama:1992gi} packages. The
adaptative Monte Carlo program VEGAS \cite{Lepage:1980dq} was employed
to perform the phase space integration.

%%%%%%%%%%%%%%%%%%%%%%%%%%%%%%%%%%%%%%%%%%%%%
\begin{table} [b]
\begin{tabular*}{\textwidth}{@{\extracolsep{\fill}}ccc}
\hline\hline
Direct & Once Resolved & Twice Resolved \\ \hline
$\gamma\gamma\to c\bar{c}g$&
$\gamma q (\bar q)\to c\bar{c}q (\bar q)$&
$q\bar{q}\to c\bar{c}g$ \\
&$\gamma g\to c\bar{c}g$&$gq(\bar q)\to c\bar{c}q(\bar q)$ \\
&&$gg\to c\bar{c}g$\\
\hline\hline
\end{tabular*}
\caption{\label{process}
Subprocesses contributing to $J/\psi$ production in $\gamma \gamma$
collisions. Here $q$  stands for the light quark flavors $ u , d , s$.}
\end{table} 
%%%%%%%%%%%%%%%%%%%%%%%%%%%%%%%%%%%%%%%%%%%%%

In the framework of the CEM, the evaluation of the photon--photon production
cross section contains only the free parameters appearing in the perturbative
QCD calculation of the subprocesses presented in Table \ref{process}, since
the CEM free parameter $\rho_\psi$ can be fixed at the value extracted from
the photo-production of $J/\psi$, {\em i.e.}  $\rho_\psi=0.5$
\cite{Amundson:1996qr}. We used the leading order GRV-G \cite{GRV} and GRS-G
\cite{Gluck:1994tv} parton density functions as provided by CERN PDFLIB
package with the partonic subprocess center--of--mass energy as factorization
scale $\mu_F=\sqrt{\hat{s}}$. We verified that our predictions do not vary
significantly for other choices of the factorization scales, {\em e.g.}
$\mu_F= \frac{1}{2}~\sqrt{\hat{s}}$ and $\mu_F=2~\sqrt{\hat{s}}$. We also
verified that the results are very similar for the GRV-G and GRS-G parton
distributions (see Table \ref{xsections}). The strong coupling constant was
evolved in leading order considering four active flavors and
$\Lambda_{QCD}^{(4)}=300$ MeV, while the charm quark mass was varied between
1.2 and 1.4 GeV.

%%%%%%%%%%%%%%%%%%%%%%%%%
\begin{figure}
\epsfig{file=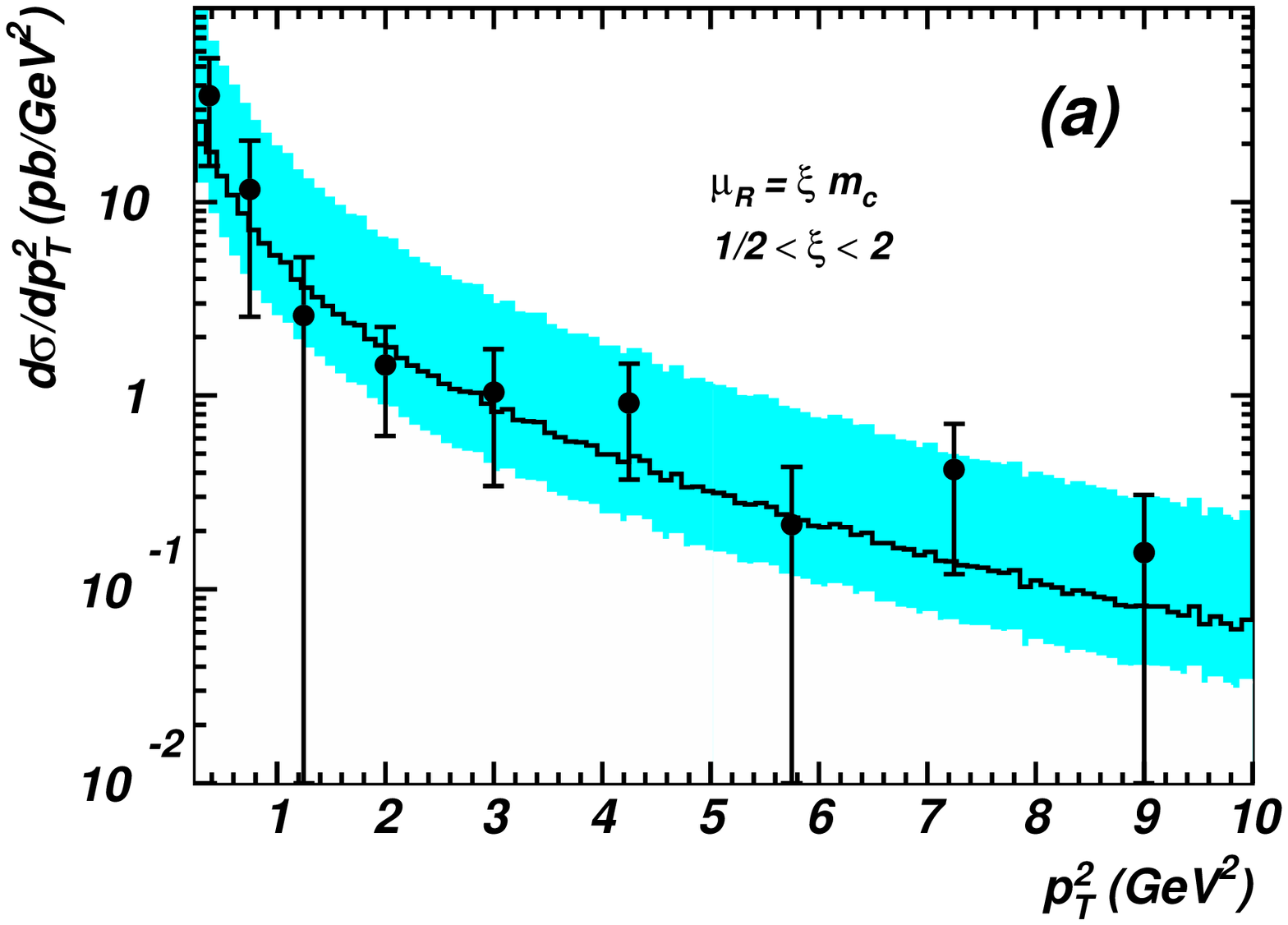,width=0.45\linewidth,clip=}
\hfill
\epsfig{file=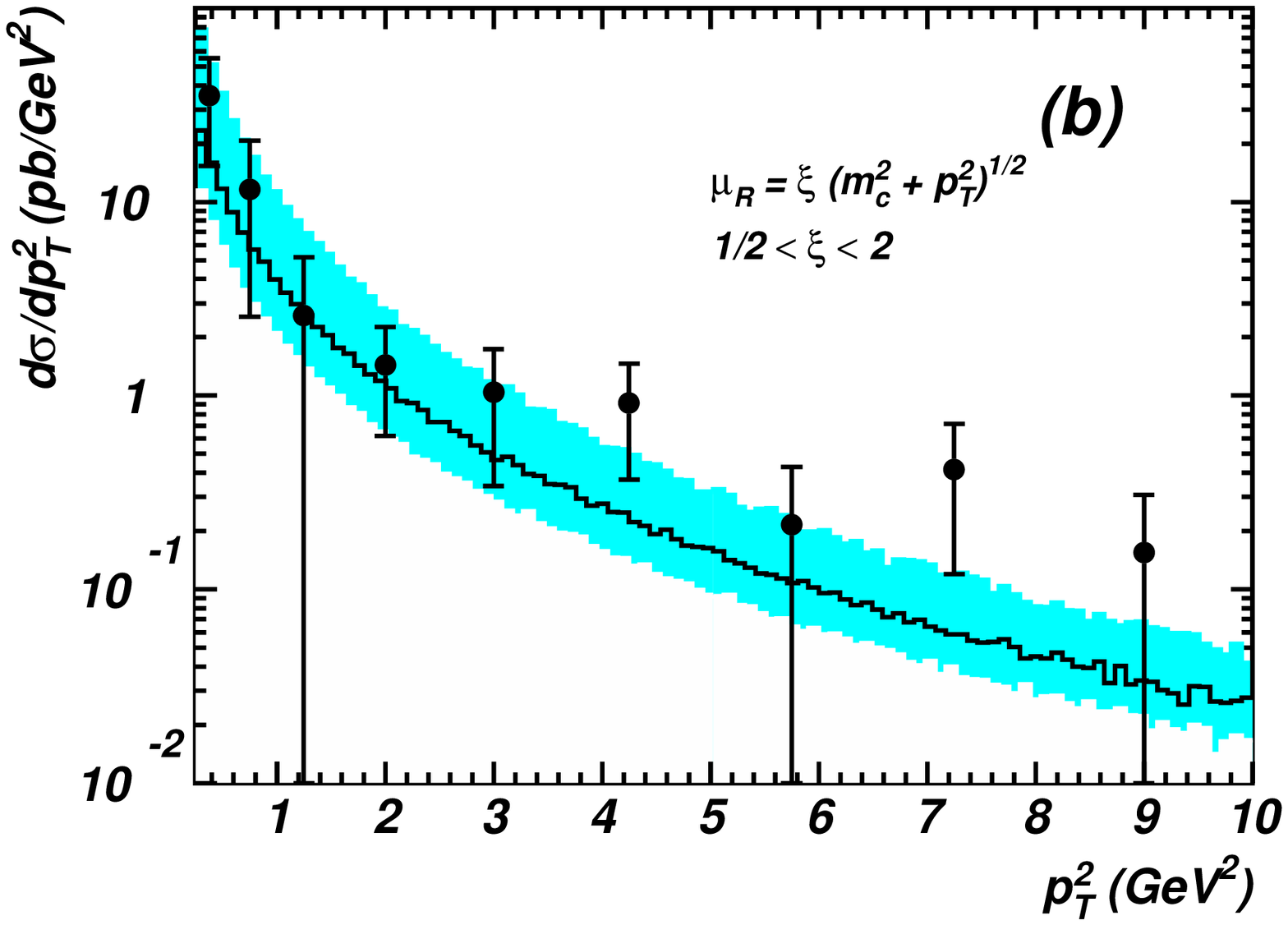,width=0.45\linewidth,clip=}
\caption{\label{scales}
Uncertainty on the $p_T^2$ differential cross section originated from
different choices of the renormalization scale $\mu_R$. In (a) we chose
$\mu_R = \xi m_c$ while in (b) $ \mu_R = \xi \sqrt{m_c^2 + p_T^2}$. The
shaded band was obtained by varying $\frac{1}{2} < \xi < 2$. We fixed
$m_c=1.3$ GeV, and used the GRS-G parton density function in both figures.}
\end{figure}
%%%%%%%%%%%%%%%%%%%%%%%%%

In order to access the theoretical uncertainties in the lowest order CEM
calculations, we analyzed the predicted $J/\psi$ transverse momentum spectrum
for different choices of the renormalization scale ($\mu_R$).  We present in
Fig.\ \ref{scales}a the predicted $p_T^2$ spectrum obtained for $\mu_R = \xi
m_c$ with $ \frac{1}{2} < \xi < 2$ and $m_c=1.3$ GeV, as well as the DELPHI
experimental results \cite{Todorova-Nova:2001pt,Klasen:2001cu}. We can see
from this figure that CEM describes well the shape of the distribution,
despite the large uncertainty in the absolute value of the differential cross
section.  Notice that we are only changing a global factor ($\alpha_S$) for
this choice of $\mu_R$ when we vary $\xi$. Figure \ref{scales}b displays the
$p_T^2$ spectrum for $\mu_R = \xi \sqrt{m_c^2 + p_T^2}$ with $\frac{1}{2} <
\xi < 2$ and $m_c=1.3$ GeV. For this choice of $\mu_R$ the uncertainties in
the the $p_T^2$ distribution are smaller than for the previous choice of
$\mu_R$. However, the shape of the $p_T^2$ spectrum changes and the CEM
prediction seems to diminish faster at large $p_T^2$ than the data.

%%%%%%%%%%%%%%%%%%%%%%%%%
\begin{figure}
\begin{center}
\epsfig{file=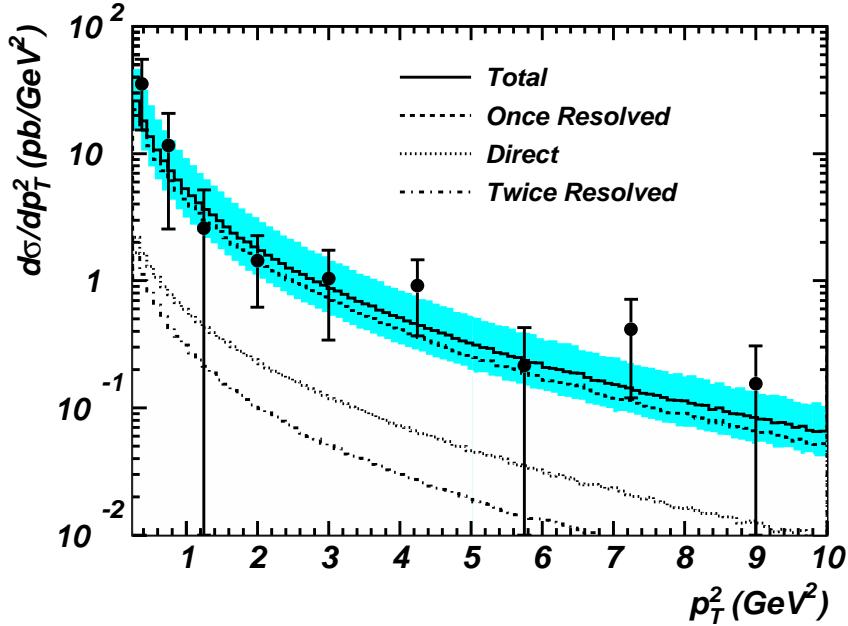,width=0.7\linewidth,clip=}
\end{center}
\caption{\label{mass}
Differential cross section as function of the squared transverse
momentum of the $J/\psi$.  The shaded band shows the theoretical
prediction obtained by varying the charm mass ($m_c = 1.3 \pm 0.1$
GeV). We explicitly show the contributions from direct, once resolved
and twice resolved cross sections for $m_c=1.3$ GeV.}
\end{figure}
%%%%%%%%%%%%%%%%%%%%%%%%%

In Figure \ref{mass} we display the contributions to the $J/\psi$ $p_T^2$
spectrum arising from direct, once resolved, and twice resolved processes.
These distributions were obtained using the GRS-G photon parton densities,
$\mu_R = m_c$ and $m_c = 1.3$ GeV. As we can see, the once resolved processes
are responsible for the majority of the events ($\simeq 85$\%) while direct
and twice resolved processes account for less than 15\% of the total cross
section. The most important process is $\gamma g \to c \bar{c} g$.  We also
present in this figure the uncertainties associated to the charm quark mass;
the shaded band represents the sum of all contributions taking $m_c=1.3\pm
0.1$ GeV. Notice that the largest uncertainties in the CEM prediction
originates from the choice of the renormalization scale. This is quite
expected since we are performing our calculation in lowest-order perturbative
QCD. We summarize our results for the total cross section in Table
\ref{xsections}.

\begin{figure}
\begin{center}
\epsfig{file=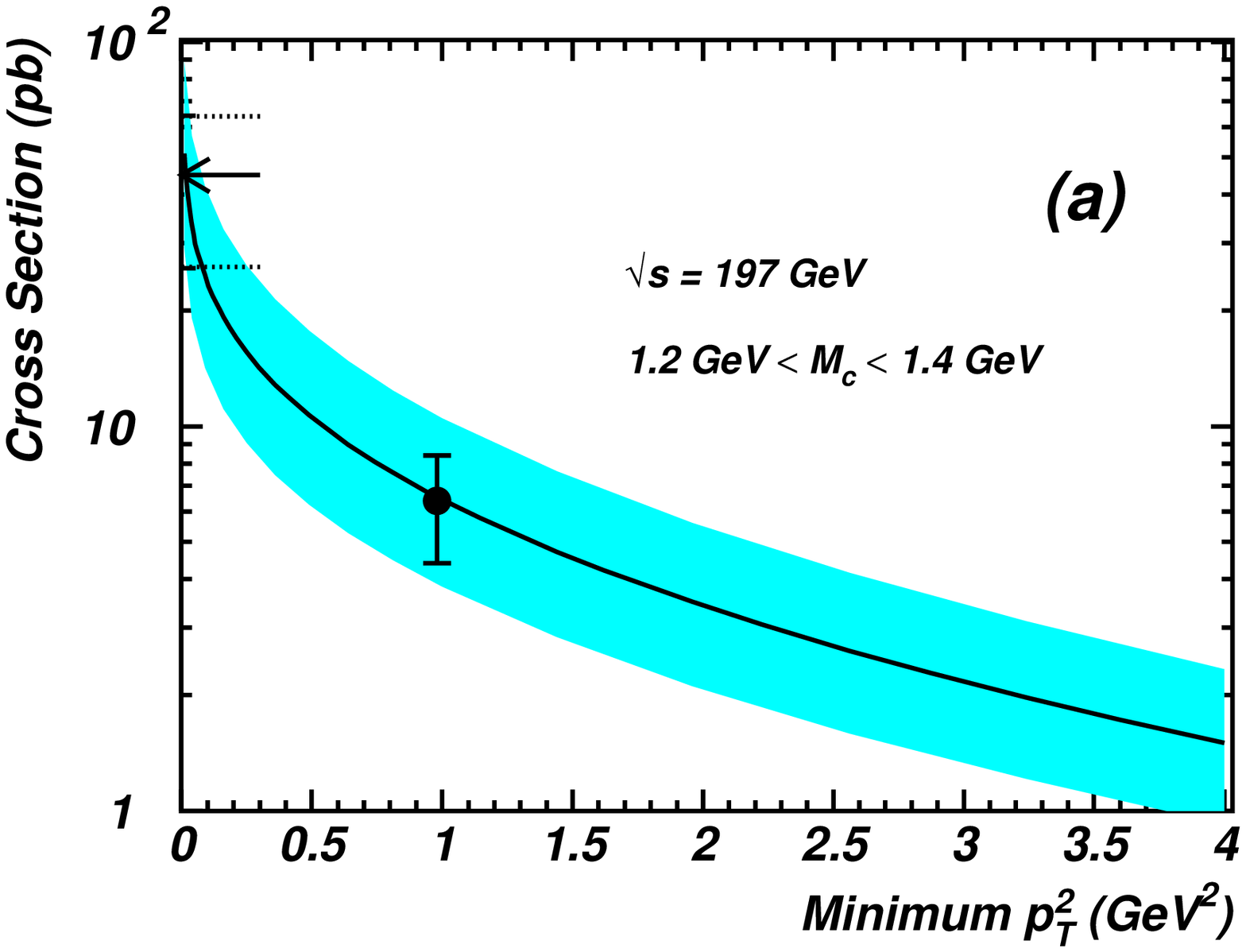,width=0.45\linewidth,clip=}
\hfill
\epsfig{file=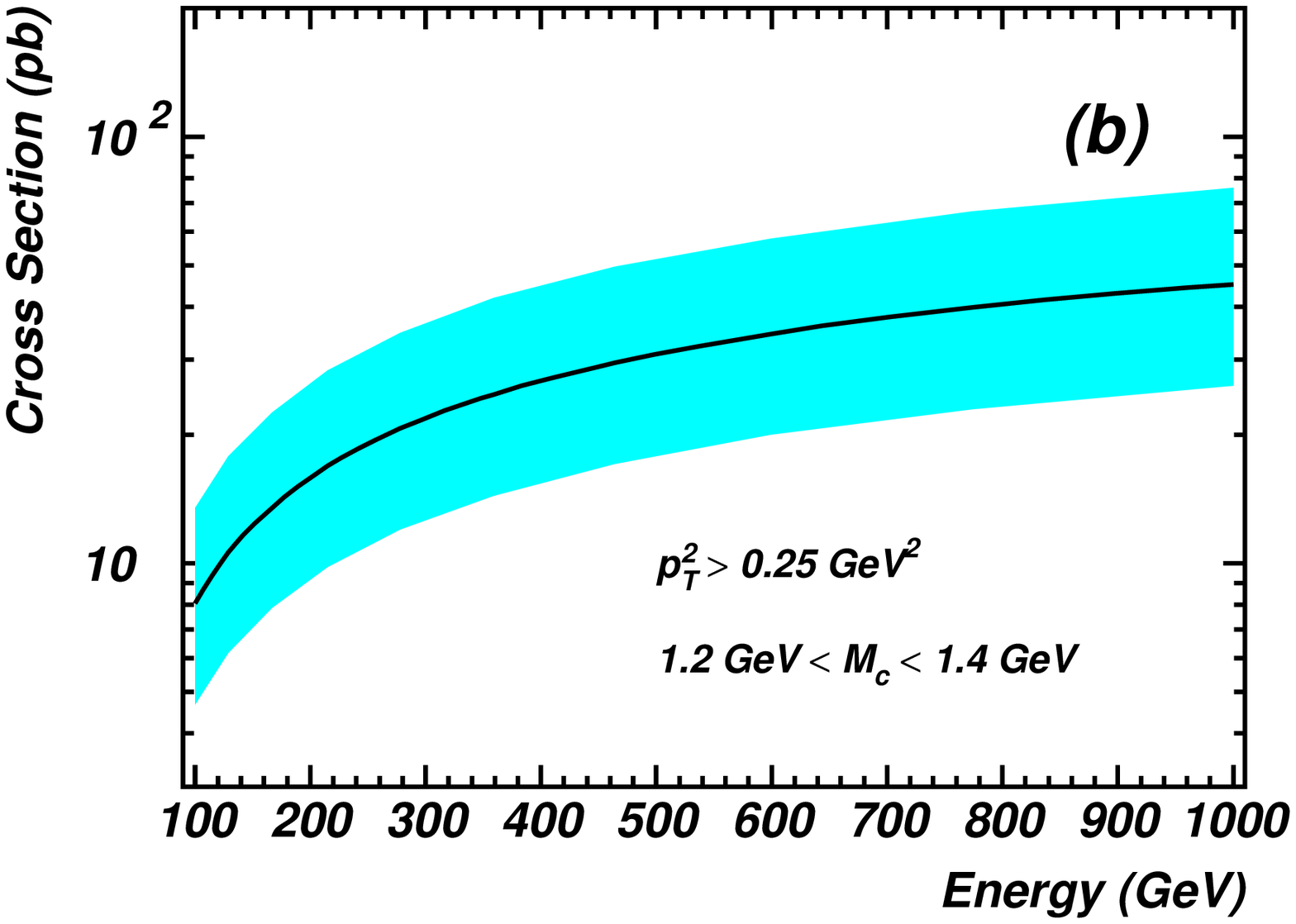,width=0.45\linewidth,clip=}
\end{center}
\caption{
  Total cross section as function of the minimum squared transverse momentum
  (a) and the $e^+e^-$ center--of--mass energy (b).  In (a) the solid line
  stands for the DELPHI measured total cross section while the dotted lines
  indicate the experimental error of this quantity. We varied the charm quark
  mass as $m_c = 1.3 \pm 0.1$ GeV to estimate the theoretical uncertainties.
  We used $\sqrt{s}= 197$ GeV for the minimum transverse momentum dependence
  (a) and imposed $p_T^2 > 0.25$ GeV$^2$ for the center--of--mass energy
  dependence (b).  The remaining parameters are the same as for Figure
  \ref{mass}.  }
\label{xsec}
\end{figure}

In order to further compare our results with the recently published DELPHI
results \cite{Abdallah:2003du,Klasen:2001cu}, we evaluated the dependence of
the total $J/\psi$ yield on the minimum transverse momentum for $\sqrt{s}=197$
GeV. The result is presented in Fig.~\ref{xsec}a.  As can be seen from this
figure, the choice of QCD parameters we used in this analysis provides a very
good description of the existing data, reinforcing our confidence on the
predictive power of of the color evaporation model.  Figure \ref{xsec}b
displays the CEM predictions for the $J/\psi$ production cross section as a
function of the $e^+ e^-$ center--of--mass energy ($\sqrt{s}$). Here we
assumed that $p_T^2 >$ 0.25 GeV$^2$, $\mu_R = m_c$ with $m_c = 1.3 \pm 0.1$
GeV, and we used the GRS-G set of parton distribution functions. As expected,
the total cross section grows with the center--of--mass energy due to the
increase in the photon-photon luminosity. We verified that contributions of
direct, once resolved and twice resolved processes are in the same proportion
of the results presented for $\sqrt{s} = 197$ GeV; see Fig.\ \ref{mass}.
Taking into account the planned luminosity of the future $e^+ e^-$ colliders,
we can easily foresee that it will be possible to extract very precise data on
the photon--photon charmonium production in these machines.

%%%%%%%%%%%%%%%%%%%%%%%%%
%%%%%%%%%%%%%%%%%%%%%%%%%

%%%%%%%%%%%%%%%%%%%%%%%%%%%%%%%%%%%%%%%%%%%%%%%%%%%%%%%%%%%%%%%%%%%%%%
\section{Conclusion}

In this paper we showed that the Color Evaporation Model for quarkonium
production correctly describes DELPHI data on $J/\psi$ via photon-photon
collisions. Due to the rather large uncertainties in the data, its is not
possible to use them to discriminate between the different proposed mechanisms
for charmonium production. As far as the DELPHI data are considered, the NRQCD
\cite{Klasen:2001cu} and CEM frameworks present equivalent results.

Considering that the CEM is also successful in describing the photo- and
hadro-production of charmonium, we conclude that this model gives a robust and
simple parameterization of all charmonium physics. Moreover, $\gamma \gamma$
reactions provide a clear proof of the importance of colored $c \bar{c}$ pairs
to the production of charmonium, since the data on this reaction can only be
explained considering resolved photon processes, which lead to colored $c
\bar{c}$ pairs.

%%%%%%%%%%%%%%%%%%%%%%%%%
\begin{table}
\setlength{\tabcolsep}{10pt}
\begin{tabular*}{\textwidth}{@{\extracolsep{\fill}}cccccccc}
\hline\hline
\multicolumn{4}{c}{Parameters}&\multicolumn{4}{c}{Cross Sections (pb)}\\
\hline
PDF&$m_c$&$\xi$&$\beta$&Direct&Once Resolved&Twice Resolved&Total\\ 
\hline
GRV-G&1.3&1.0&0&1.72 &13.3  &1.00 &16.1\\
GRS-G&1.3&1.0&0&1.72 &13.0  &0.94 &15.7\\
GRS-G&1.2&1.0&0&2.75 &21.8  &1.73 &26.3\\
GRS-G&1.4&1.0&0&1.02 & 7.5  &0.51 & 9.07\\
GRS-G&1.3&0.5&0&3.26 &46.8  &6.42 &56.4\\
GRS-G&1.3&0.5&1&2.42 &28.4  &3.15 &34.0\\
GRS-G&1.3&1.0&1&1.44 & 9.65 &0.61 &11.7\\
GRS-G&1.3&2.0&0&1.17 & 5.99 &0.29 & 7.45\\
GRS-G&1.3&2.0&1&1.03 & 4.84 &0.22 & 6.08\\
\hline\hline
\end{tabular*}
\caption{\label{xsections}
Cross sections for direct, once resolved, and twice resolved production
processes for $p_T^2>0.25$ GeV$^2$ using different sets of parton
distribution functions, charm masses, and renormalization scales $\mu_R
= \xi\sqrt{m_c^2 + \beta p_T^2}$.} 
\end{table}
%%%%%%%%%%%%%%%%%%%%%%%%%

%%%%%%%%%%%%%%%%%%%%%%%%%%%%%%%%%%%%%%%%%%%%%%%%%%%%%%%%%%%%%%%%%%%%%%
\begin{acknowledgments}
  This research was supported in part by Funda\c{c}\~{a}o de Amparo \`a
  Pesquisa do Estado de S\~ao Paulo (FAPESP), by Conselho Nacional de
  Desenvolvimento Cient\'{\i}fico e Tecnol\'ogico (CNPq), and by Programa de
  Apoio a N\'ucleos de Excel\^encia (PRONEX).
\end{acknowledgments}

%%%%%%%%%%%%%%%%%%%%%%%%%%%%%%%%%%%%%%%%%%%%%%%%%%%%%%%%%%%%%%%%%%%%%%

\end{document}